\documentclass{moriond}

\usepackage{amsmath,amssymb}

\def\be{\begin{equation}}
\def\ee{\end{equation}}
\def\bea{\begin{eqnarray}}
\def\eea{\end{eqnarray}}

\newcommand{\Photo}{}

\begin{document}
\vspace*{4cm}
\title{IS THE STANDARD MODEL SCALAR THE FIRST DISCOVERED SUSY PARTICLE?}

\author{ C. BIGGIO }

\address{Dipartimento di Fisica, Universit\`a di Genova \& INFN,
  Sezione di Genova, \\
via Dodecaneso 33, 16159 Genova, Italy}

\maketitle
\abstracts{The scalar particle recently discovered at the LHC has the same gauge
quantum numbers as the neutrino, so they could be one the superpartner
of the other. We discuss the conditions that should be satisfied in
order to realize such identification and present a model where this is
realized. This model possesses an interesting phenomenology that we
present here.}

\section{Introduction}

On July 4$^{th}$ 2012, at CERN, it was announced the discovery of a
new particle, at a mass of $(125.9\pm 0.4)$~GeV, compatible with the
Standard Model (SM) scalar boson. In October 2013 the Nobel Prize for
physics was assigned to Fran\c{c}ois Englert and Peter W. Higgs ``for the
theoretical discovery of a mechanism that contributes to our
understanding of the origin of mass of subatomic particles, and which
recently was confirmed through the discovery of the predicted
fundamental particle, by the ATLAS and CMS experiments at CERN's Large
Hadron Collider''.  This discovery completes the picture of the SM
particles and, for the first time, a scalar is discovered with the
same gauge quantum numbers of an already discovered fermion, the
neutrino. Indeed in the SM both the scalar and the neutrinos belong to
SU(2) doublets and have the same hypercharge. Then a natural question
arises, if they can be one the superpartner of the other. Were this
the case, the SM scalar discovery would also represents the discovery
of supersymmetry (SUSY). It is then clear that it is worth to check
whether this is feasible or not. This has been studied in details in
Ref.~\cite{Riva:2012hz},
of which in the following I will give a short review.\\

In order to identify the SM scalar with a sneutrino few
conditions have to be satisfied. Since we are dealing with a SUSY
theory, we could think of starting from the Minimal
Supersymmetric Standard Model (MSSM). In order to avoid fast proton
decay, in the MSSM a parity symmetry, called R-parity, is usually imposed, under which all SM
particles are even, while SUSY partners are odd. Additionally, this ensures
the absence of large neutrino masses, that would be generated if the
R-parity were broken. If we now identify
a sneutrino with the SM scalar, once the latter acquires vev, the
R-parity and Lepton Number (LN), under which the sneutrino is charged, will be broken and large neutrino
masses will be generated, thus excluding the model. We therefore
immediately come to the first condition: in order to identify the SM scalar
with a sneutrino \textit{the R-parity should be replaced by a $U(1)_R$ symmetry,
which plays the r\^ole of LN and prevents the generation of large
$\nu$ masses}. 

$U(1)_R$ symmetries are well-known in the literature, since they have
the interesting feature that, in models respecting them, gauginos have
Dirac masses, and this offers a further protection mechanism to the SM
scalar mass known as ``supersoft SUSY breaking''~\cite{Fox:2002bu}. In
this framework it is therefore possibile to identify a slepton doublet
with the SM scalar doublet and indeed this possibility has been
explored in the literature~\cite{Hsnu}. However, in order to obtain
all the SM Yukawa Lagrangian terms and therefore all the fermion
masses, a second chiral superfield, $H_u$ is usually introduced, since
a term containing the hermitian conjugate of the SM scalar, like in
the up-quark Yukawa coupling in the SM, is forbidden in SUSY theories,
since the superpotential must be analytic. Then, in the previously
mentioned models, the sneutrino plays the r\^ole of the SM scalar only
in the down quark and charged lepton sectors, while $H_u$ provides the
masses of the up-type quarks. Once $H_u$ is introduced, a new
superfield has to be added to cancel the anomaly induced by the
presence of $H_u$ alone. Moreover, the $\mu$-term is still there in
this case, with all the problematics associated to it.

Here we would like to explore the viability of a model where no chiral
superfield is introduced, a part from the ones associated to leptons
and quarks, and where a sneutrino plays the r\^ole of the SM scalar,
giving mass to $all$ the fermions. This would have the additional
benefits that no superfield must be added in order to cancel any
anomaly and moreover the $\mu$-problem would be solved, since no
$\mu$-term would be present in the theory. The price one has to pay in
order to realize this is that \textit{at least the Yukawa coupling of
  up-type quarks must come from the Kh\"{a}ler potential, i.e. from
  higher-dimensional SUSY-breaking operators}.

In the following I will discuss the main features of the model and the
interesting phenomenology associated to it.

\section{The Higgsinoless MSSM}

\begin{table}[t]
\caption{Superfield  content  and charge assignments under the SM gauge group and the $U(1)_R$ symmetry.}
\label{table1}
\vspace{0.4cm}
\begin{center}
\begin{tabular}{|c|c|c|}
\hline
& 
\textbf{$  SU(3)_c\times SU(2)_L\times U(1)_Y $}& 
\textbf{$U(1)_R$} \\ 
\hline 
$Q$ & \  \ \ \ \ \ \ $ (3,2)_{\frac{1}{6}}$ \ \ & $1+B $\\ 
$U$ & \  \ \ \ \ \ \ $ (\bar 3,1)_{-\frac{2}{3}}$ \ \ & $1-B$ \\  
$D$ & \  \ \ \ \ \ \ $ (\bar 3,1)_{\frac{1}{3}}$ \ \ & $1-B$ \\ 
$L_{1,2}$ & \  \ \ \ \ \ \ $ (1,2)_{-\frac{1}{2}}$ \ \ & $1-L$ \\ 
$E_{1,2}$ & \  \ \ \ \ \ \ $ (1,1)_{1}$ \ \ & $1+L $\\ 
$H\equiv L_3$ & \  \ \ \ \ \ \ $ (1,2)_{-\frac{1}{2}}$ \ \ & 0 \\
$E_3$ & \  \ \ \ \ \ \ $ (1,1)_{1}$ \ \ & 2 \\  
$W^\alpha_a$ & \  \ \ \ \ \ \ $ (8,1)_0+(1,3)_0+(1,1)_0 $\ \ & 1 \\ 
\hline
$ \Phi_a $ &  \  \ \ \ \ \ \ $ (8,1)_0+(1,3)_0+(1,1)_0$ & 0 \\ 
\hline
$ X\equiv \theta^2F $ &  \  \ \ \ \ \ \ $  (1,1)_0$ & 2\\ 
\hline
\end{tabular} 
\end{center}
\end{table} 

Since the only chiral superfields present in this model are the ones
associated to quarks and leptons, no chiral Higgs superfields are
present and hence no Higgsinos, from which the name of the model. 
In table~\ref{table1} the superfield content of this model, as well
as their charge assignments, are shown. From there one can see that a certain
freedom on the charges is possibile. In the following we will not enter
into the details of such charge assignment, for which we remand to
Ref.~\cite{Riva:2012hz}, but just assume that $L\ne
1$ (i.e. the SM scalar is identified with $L_3$) and $B\ne  1/3$ (in
order to prevent proton decay). It follows that the only possible superpotential
terms, at the renormalizable level, are:
\begin{equation}\label{W}
W= Y_d\, HQ  D+Y_{e\, ij}\, HL_i E_j\, ,
\end{equation}
where indexes $i,j=1,2$ are summed over ($i,j=3$ is forbidden by the
antisymmetry of SU(2) contracted indexes) and $Y_d$ is a matrix in
flavor space.  As it stands, this superpotential does not generate
up-type quark masses, gaugino masses, nor a mass for the charged
lepton contained in $L_3$, whose scalar partner plays the r\^ole of
the SM scalar. These must originate as SUSY breaking effects.  If we
introduce a spurion field $X$, whose $F$-component is nonzero,
$X=\theta^2 F$, and breaks SUSY, in a SUSY-preserving notation the
masses of the up-type quarks can be written as
\begin{equation}
\label{topmass}
\int d^4\theta\ y_u\frac{  X^\dagger}{M} \frac{  H^\dagger Q U}{\Lambda}  = \int d^2\theta\ Y_uH^\dagger Q U\, ,
\end{equation}
where $y_u$ are dimensionless couplings and $Y_u\equiv
y_uF/(M\Lambda)$ are the Yukawa couplings of the up-type quarks. Here
$\Lambda$ is the scale at which the effective operator is generated,
while $M$ is the SUSY mediation scale.

In a similar fashion we can write the masses for $\ell_3$ and the
gauginos
\begin{eqnarray}
\label{masses}
 &&\int d^4\theta\ y_3\frac{X^\dagger  X}{M^2} \frac{H D^\alpha H D_\alpha E_3}{\Lambda^2} \\
&& \int d^2\theta\ \frac{D^\alpha X}{M}W^{a}_\alpha \Phi_{a} \, ,
\end{eqnarray}
where $D^\alpha$ are superspace derivatives,
as well as an additional quartic coupling for the SM scalar:
\begin{equation}
\int d^4\theta\  \lambda_H \frac{X^\dagger X}{M^2}\frac{|H|^4}{\Lambda^2}= \delta \lambda_h\, h^4+\dots  \, .
\label{extraquartic}
\end{equation}
This last term is needed in order to obtain a SM scalar mass $\sim 126$
GeV, since in this model A-terms are forbidden by the R-symmetry and,
for naturalness, we would like to have light stops.

Eqs.~(\ref{W})-(\ref{extraquartic}) are the necessary and
sufficient ingredients we need to build a realistic model without any
additional chiral superfield. Moreover, from Eq.~(\ref{topmass}) we
derive the first interesting consequence of our approach: since
$Y_t\sim 1$, and $F/M$ is the scale of scalar superpartners that we
would like not to be heavier than the TeV, we obtain 
$\Lambda\sim y_u F/M\lesssim 4\pi\,   {\rm TeV}$. That is, this model
is \textit{an effective theory valid up to few tens of TeV}.

A nice feature of this model is that, thanks to the R-symmetry, the
SUSY breaking corrections to the SM scalar mass are suppressed, and
EWSB is realized without fine tuning. Moreover, for the same reason, a
naturally splitted spectrum is possible. In particular, gauginos must
be heavier then the TeV, due to the modifications they can induce in
the $Z$ coupling to charged leptons, with whom they mix, while stops
and sbottoms can be just around the corner in LHC searches. We will
not enter here into these technical details, that can be found in
Ref.~\cite{Riva:2012hz}, while we will now focus on the interesting
phenomenology proper of this model.

\section{Phenomenology}

Let's start with the SM scalar. At tree level, its couplings are
identical to the SM ones, but at the loop level deviations can
occur. In particular, light stops circulating in the loops can induce
corrections to the scalar couplings to gluons and photons,
modifying the branching ratios and especially the production cross
section. 

An even more interesting feature is given by the possibility of having
an invisible branching fraction. Indeed, since the scalar is the
partner of the neutrino and they couple to the goldstino, if the
gravitino is light, the scalar can decay into gravitino and neutrino,
which manifest themselves in missing transverse energy ($E_T$). For
$F\sim 1$ TeV, the
invisible branching fraction can be as large as 10\%.\\

\begin{table}[t]
\caption{Decay modes for the (third family) squarks
with the corresponding Lagrangian interaction.}
\label{table2}
\vspace{0.4cm}
\begin{minipage}[b]{0.5\linewidth}\centering
\begin{tabular}{|l|l|}
\hline
Decay  & Interaction \\
\hline
\hline
$\tilde{t}_L \to b_R \bar l^-_L $  & $Y_d\, HQD|_{\theta^2}$  \\
\hline
$\tilde{t}_L \to t_R \bar{\nu}_L $ &$\frac{1}{\Lambda^2}|H|^2|Q|^2|_{\theta^4}$ \\
\hline
$\tilde{t}_L \to t_L \tilde{G} $ &$\frac{m_{t}^2-m_{\tilde{t}_L}^2}{F}\,\tilde{t}_L^*\tilde{G}\,t_L$ \\
\hline
\hline
$\tilde{b}_L \to b_R \bar{\nu}_L $  & $Y_d\, QHD|_{\theta^2}$ \\
\hline
$\tilde{b}_L \to b_L  \tilde{G}  $   &  $ \frac{m_{b}^2-m_{\tilde{b}_L}^2}{F}\,\tilde{b}_L^*\tilde{G}\,b_L$ \\
\hline
\end{tabular}
\end{minipage}
\begin{minipage}[b]{0.5\linewidth}
\centering
\begin{tabular}{|l|l|}
\hline
Decay  & Interaction \\
\hline
$\tilde{t}_R \to t_L \nu_L $ & $\frac{1}{\Lambda^2}|H|^2|U|^2|_{\theta^4}$  \vspace{0.5mm} \\
\hline
$\tilde{t}_R \to t_R \bar{\tilde{G}} $ & $ \frac{m_{t}^2-m_{\tilde{t}_R}^2}{F}\,\tilde{t}_R^*\bar{\tilde{G}}\,\bar{t}_L$\\
\hline
\hline
$\tilde{b}_R \to b_L \nu_L $  & $Y_d\, QHD|_{\theta^2}$  \\
\hline
$\tilde{b}_R \to t_L \, l^-_L $  & $Y_d\, QHD|_{\theta^2}$  \\
\hline
$\tilde{b}_R \to b_R  \bar{\tilde{G}} $   &  $ \frac{m_{b}^2-m_{\tilde{b}_R}^2}{F}\tilde{b}_R^*\bar{\tilde{G}}\,\bar{b}_L$ \\
\hline
\end{tabular}
\end{minipage}
\end{table}

In this model squarks and sleptons can be light. In particular, for
naturalness reasons, we expect the third generation squarks to be
lighter than the TeV scale. Notice that, thanks to the R-symmetry,
left and right sfermions do not mix. This permits us to make a
nice prediction on quark masses:
\begin{equation}
\label{sbottomstop}
m^2_{\tilde{b}_L}=m^2_{\tilde{t}_L}-m^2_{t}+m^2_{b}\ .
\end{equation}
The decay modes of sfermions are dictated by symmtries and, for the
third generation squarks, are summarized in table~\ref{table2}. Notice
that in many cases stops and sbottoms decay into a quark plus a
gravitino or a neutrino, i.e. missing energy. Therefore, the MSSM
searches for a squark decaying into a quark plus a neutralino can be
adapted here by taking $m_{\tilde{\chi}^0} = 0$ and used to place bounds on
the masses of squarks. However, at difference with the MSSM and due to
the absence of the R-parity, here the squarks can decay into a quark and a
lepton. In particular the $\tilde{t}_L$ can decay into a $b$ and a charged
lepton, while the $\tilde{b}_R$ can decay into a $t$ and a $\ell$. The
branching ratios into these channels depend on several variables, in
particular on the gravitino mass: if the gravitino is light the decay
into it and a quark dominates, while if it is heavy the above
leptoquark decays can occur. As for the decay $\tilde{t}_L \to b_R
\bar l^-_L $ one can adapt leptoquark searches at the LHC and put a
bound on the stop mass. On the other hand, the decay $\tilde{b}_R \to
t_L \, l^-_L $ has not yet been searched for.

The prediction on the mass relation, the fact that leptoquark decays
exist with predictable branching ratios and that the final state helicity
is fixed in this model, render it distinguishable from the
MSSM. Indeed, suppose a final state composed by a b-jet and missing
transverse energy is observed: it can be the $\tilde{b}_R$ of this
model only if a leptoquark decay into top+lepton is observed at the
same mass, or it can be the  $\tilde{b}_L$ if the $\tilde{t}_L$ is
observed at a slightly higher mass. On the other hand suppose that a
top and missing $E_T$ are observed: it can be the $\tilde{t}_L$ if a lighter
$\tilde{b}_L$ and decays into b and leptons are observed, but it can
be also the $\tilde{t}_R$. In this case, in order to distinguish from
the MSSM one should look at the top helicity and, even if not trivial,
this is in principle feasible. \\

\begin{table}[t]
\caption{Dominant decay modes for first and second family squarks
when the  gravitino is heavy or $\sqrt{F}\gg$ TeV.}
\label{table3}
\vspace{.4cm}
\centering
\begin{tabular}{|l||l||l|}
\hline
$\tilde{u}_L \to d +\bar l^-_L+Z $ & $\tilde{c}_L \to s +\bar
l^-_L\,\,\,\,\,  (\text{for}\ m_{\tilde{c}_L}\lesssim500~\textrm{GeV})$ &
$\tilde{c}_R \to c+ \nu_L \,\,\,\,\, (\text{for}\
m_{\tilde{c}_R}\lesssim 600~\textrm{GeV})$\\\cline{1-1}
$\tilde{d}_L \to u+\bar \nu_L+W^- $& $\hspace{5.3mm} \to s+\bar l^-_L +Z$ &$\hspace{5.3mm} \to c+ l^-_L +W^+$ \\
\hline
$\tilde{u}_R \to u+ l^-_L+W^+$ &$\tilde{s}_L \to s+\bar
\nu_L$$\,\,\,\,\, (\text{for}\ m_{\tilde{s}_L}\lesssim 300~\textrm{GeV})$&$\tilde{s}_R \to c+ l^-_L $ (50\%)\\
\cline{1-1}
$\tilde{d}_R \to d+ l^-_L+W^+ $&$\hspace{5.3mm} \to c+\bar \nu_L +W^-$&$\hspace{5.3mm} \to s+ \nu_L $  (50\%)\\
\hline
%$\tilde{d}_L \to d + \tilde{G}  $   &  $- \frac{m_{\tilde{d}_L}^2}{F}\,\tilde{d}_L^*\tilde{G}\,b_L$ \\
%\hline
\end{tabular}
\end{table}

\begin{table}[t]
\caption{Dominant decay modes for sleptons when the  gravitino is heavy or $\sqrt{F}\gg$ TeV.
We assume  that the slepton masses are larger than 500 GeV.}
\label{table4}
\vspace{.4cm}
\centering
\begin{tabular}{|l||l||l|}
\hline
$\tilde{e}_L \to \nu_e +\bar \nu_L+W^- $&$\tilde{\mu}_L \to \nu_\mu + \bar{\nu}_L+ W^-$ &  $\tilde{\tau}_L \to \tau+\bar \nu_L$\\
\hline
$\tilde{e}_R \to e + l^-_L+W^+ $& $\tilde{\mu}_R \to \mu +\nu_L$ (50\%)&$\tilde{\tau}_R \to \tau +\nu_L$  (50\%) \\
&$\hspace{5.3mm}  \to \nu_\mu + l^-_L$  (50\%)&$\hspace{5.4mm}  \to  \nu_\tau + l^-_L$  (50\%)\\
\hline
$\tilde{\nu}_e \to e +\bar l^-_L+Z $& $\tilde{\nu}_\mu  \to \mu+ Z +\bar{l}^-_L$&$\tilde{\nu}_\tau \to \tau +\bar l^-_L$ \\
\hline
\end{tabular}
\end{table}

As for the 1$^{st}$ and 2$^{nd}$ generation squarks, bounds coming
from the searches of final states with jets and missing $E_T$ are quite strong, namely
$>830$~GeV. In principle also for these sparticles leptoquark
decays can occur, like for stops and sbottoms, but, since they come
from superpotential terms, the corresponding branching ratios are
proportional to the Yukawa couplings that, in this case, are
small. Therefore, an interesting thing can happen: 3-body decays can
be dominant over the 2-body ones. In table~\ref{table3} the possible
decays are shown. These constitute another interesting and peculiar
signal of this model.\\

In the slepton sector the situation is quite similar: indeed also in
this case the Yukawa couplings are small and it can happen that 3-body decays
dominate. The possible decays for the charged sleptons, in the case
where the gravitino is heavy, are shown in table~\ref{table4}.\\

We have discussed here the phenomenology of a model which shows
peculiar features with respect to the MSSM. In particular we stress
that this model is testable at the LHC, distinguishable from the
MSSM in case of a discovery, and moreover falsifiable at the LHC
running at
14~TeV, since, being it an effective theory valid up to few TeV,
almost all its parameter space will be explored in the next LHC run.

\section{Conclusions}

We have answered the question of the title and shown that it is indeed
possible for the SM scalar to be the
superpartner of the neutrino. For this to be realized, few conditions
have to be satisfied, which lead to a model which is an effective
theory valid up to few TeVs. This model has an interesting collider
phenomenology that permits to test it at the LHC and, in case SUSY is
discovered, to distinguish it from the MSSM or other SUSY models.

\section*{Acknowledgments}

I would like to thank the organizers of the conference for the
pleasant and stimulating athmosphere. This work is partially supported
by the Marie Curie CIG program, project number PCIG13-GA-2013-618439. 

%\section*{Appendix}

\end{document}